\documentclass[proceedings]{JHEP3}
\usepackage{amsfonts}
\usepackage{amsmath}
\usepackage{epsfig,multicol}

\newbox\mybox

\newcommand\fverb{\setbox\mybox=\hbox\bgroup\verb}
\newcommand\fverbdo{\egroup\medskip\noindent\fbox{\unhbox\mybox}\ }
\newcommand\fverbit{\egroup\item[\fbox{\unhbox\mybox}]}
\conference{Minimal areas}
\abstract{We demonstrate that dynamical noncommutative space-time will give rise to deformed oscillator algebras. In turn, starting from some q-deformations of these algebras in a two dimensional space for which the entire deformed Fock space can be constructed explicitly, we derive the commutation relations for the dynamical variables in noncommutative space-time. We compute minimal areas resulting from these relations, i.e. finitely extended regions for which it is impossible to resolve any substructure in form of measurable knowledge. The size of the
regions we find is determined by the noncommutative constant and
the deformation parameter q. Any object in this type of space-time structure has to be of membrane type or in certain limits of string type.}

\title{Minimal areas from q-deformed oscillator algebras}
\author{Andreas Fring$^\bullet$, Laure Gouba$^\circ$ and Bijan Bagchi$^*$ \\
$^\bullet$ Centre for Mathematical Science, City University London,\\
$\,\,$ Northampton Square, London EC1V 0HB, UK\\
$^\circ$ National Institute for Theoretical Physics (NITheP), Stellenbosch
7600, South Africa \\
$^*$ Department of Applied Mathematics, University of Calcutta,\\
$\,\,$ 92 Acharya Prafulla Chandra Road, Kolkata 700 009, India\\
E-mail: a.fring@city.ac.uk, gouba@sun.ac.za,bbagchi123@rediffmail.com}

\input{tcilatex}

\begin{document}

\section{Introduction}

The idea to extend the quantization procedure from canonical variables to
space-time itself \cite{Snyder:1946qz} traces back over sixty years. In
recent years this general possibility has become more and more appealing,
especially in the context of quantum field theories as such type of
space-time structures will introduce natural cut-offs and theories on them
are therefore renormalized by construction \cite{Douglas:2001ba,Szabo:2001kg}%
. In addition, almost all possible theories of quantum gravity require
non-Minkowskian space-time in one form or another \cite%
{calmet,String1,String2,String3,Ashtekar}.

One of the interesting consequences of these type of space-time structures
is that in many cases they lead to modifications of Heisenberg's uncertainty
relations, which in turn result in the emergence of minimal lengths. This
means in such spaces one has almost inevitably definite fundamental
distances below which no substructure can be resolved \cite%
{Kempf1,Kempf2,Brodimas,Bieden,MacF,ChangPu,QuesneTK,AFBB,Hossenfelder}.
Recently some of us proposed \cite{AFLG} a consistent dynamical
noncommutative space-time structure in a two dimensional space which leads
to a fundamental length in one direction, implying that objects in these
spaces are of string type. Here we provide a different type of dynamical
noncommutative space-time implying a fundamental length in each of the two
directions, thus giving rise to minimal areas for which any substructures is
beyond measurable knowledge. In our construction procedure we will not only
postulate the deformed Heisenberg canonical commutation relations and check
their consistency, but we will also derive them from some more extensively
studied and more fundamental structure, namely q-deformed oscillator
algebras for which the entire Fock space can be constructed explicitly \cite%
{Bieden,MacF,ChangPu}.

In section 2 we commence with various consistent deformations of
Heisenberg's canonical commutation relations and investigate the
consequences on the commutation relations of the associated oscillator
algebra. We find that the latter are almost inevitably deformed. In section
3 we take this fact into account and reverse the setting by starting instead
from a well suited q-deformed oscillator algebra and derive from it
Heisenberg's uncertainty relations for the dynamical variables. In section 4
we briefly recall the standard argument leading to minimal length and
compute the minimal area for a selected algebra. Our conclusions and an
outlook to further open problems are stated in section 5.

\section{Creation and annihilation operators from noncommutative space-time}

\subsection{Oscillator algebras in flat noncommutative space-time}

Noncommutative flat space-time in two dimensions manifests itself in the
following modification of Heisenberg's canonical commutation relations for
the dynamical variables 
\begin{equation}
\begin{array}{lll}
\lbrack x_{0},y_{0}]=i\theta , & [x_{0},p_{x_{0}}]=i\hbar ,\qquad \quad & 
[y_{0},p_{y_{0}}]=i\hbar , \\ 
\lbrack p_{x_{0}},p_{y_{0}}]=0,\qquad \quad & [x_{0},p_{y_{0}}]=0, & 
[y_{0},p_{x_{0}}]=0.%
\end{array}
\label{1}
\end{equation}%
Restricting the noncommutative constant to be real, i.e.~$\theta \in \mathbb{%
R}$, ensures that $x_{0}$ and $y_{0}$ are Hermitian operators. We now wish
to find a representation for creation and annihilation operators in terms of
the dynamical variables $x_{0},y_{0},p_{x_{0}},p_{y_{0}}$ satisfying the
standard commutation relations for a Fock space representation 
\begin{equation}
\lbrack a_{i},a_{j}^{\dagger }]=\delta _{ij},\qquad \lbrack
a_{i},a_{j}]=0,\qquad \lbrack a_{i}^{\dagger },a_{j}^{\dagger }]=0\qquad 
\text{for }i,j=1,2.  \label{aaa}
\end{equation}%
In order to reduce the number of unknown coefficients in a possible Ansatz
for the $a_{i},a_{i}^{\dagger }$ we may take the properties of the dynamical
variables under a $\mathcal{PT}$-transformation as a guiding principle.
These type of considerations have proved to be very fruitful, allowing even
a consistent formulation of non-Hermitian systems with real eigenvalues, see
e.g.~\cite{Benderrev,Rev3,paulo} for a review or \cite{special2,specialindia}
for recent special issues. For this purpose we note that the relations (\ref%
{1}) are $\mathcal{P}_{x}\mathcal{T}$-symmetric and $\mathcal{P}_{y}\mathcal{%
T}$-symmetric in the sense that they remain invariant under a simultaneous
reflection in the $x_{0}$-direction together with a time reversal and under
a simultaneous reflection in the $y_{0}$-direction together with a time
reversal, respectively, 
\begin{equation}
\begin{array}{llllll}
\mathcal{P}_{x}\text{: \ \ \ } & x_{0}\mapsto -x_{0},~ & y_{0}\mapsto y_{0},~
& p_{x_{0}}\mapsto -p_{x_{0}},~ & p_{y_{0}}\mapsto p_{y_{0}},~ &  \\ 
\mathcal{P}_{y}\text{: \ \ \ } & x_{0}\mapsto x_{0},~ & y_{0}\mapsto -y_{0},~
& p_{x_{0}}\mapsto p_{x_{0}},~ & p_{y_{0}}\mapsto -p_{y_{0}},~ &  \\ 
\mathcal{T}\text{:} & x_{0}\mapsto x_{0},~ & y_{0}\mapsto y_{0},~ & 
p_{x_{0}}\mapsto -p_{x_{0}},~ & p_{y_{0}}\mapsto -p_{y_{0}},~ & i\mapsto -i,
\\ 
\mathcal{P}_{x}\mathcal{T}\text{:} & x_{0}\mapsto -x_{0},~ & y_{0}\mapsto
y_{0},~ & p_{x_{0}}\mapsto p_{x_{0}},~ & p_{y_{0}}\mapsto -p_{y_{0}},~ & 
i\mapsto -i, \\ 
\mathcal{P}_{y}\mathcal{T}\text{:} & x_{0}\mapsto x_{0},~ & y_{0}\mapsto
-y_{0},~ & p_{x_{0}}\mapsto -p_{x_{0}},~ & p_{y_{0}}\mapsto p_{y_{0}},~ & 
i\mapsto -i.%
\end{array}
\label{PT}
\end{equation}%
We demand now to have a definite transformation property for the $%
a_{i},a_{i}^{\dagger }$, that is we would like them to be either even or odd
under a $\mathcal{P}_{x,y}\mathcal{T}$-transformation, i.e. $a_{i}\mapsto
a_{i}$, $a_{i}^{\dagger }\mapsto a_{i}^{\dagger }$ or $a_{i}\mapsto -a_{i}$, 
$a_{i}^{\dagger }\mapsto -a_{i}^{\dagger }$, such that we can use this
property to reduce the total number of constants. Assuming that the
dependence on the $x_{0},y_{0},p_{x_{0}},p_{y_{0}}$ is still linear, the
general operators of the form 
\begin{equation}
\begin{array}{ll}
a_{1}:=\alpha _{1}x_{0}+i\alpha _{2}y_{0}+i\alpha _{3}p_{x_{0}}+\alpha
_{4}p_{y_{0}},\qquad & a_{1}^{\dagger }:=\alpha _{1}x_{0}-i\alpha
_{2}y_{0}-i\alpha _{3}p_{x_{0}}+\alpha _{4}p_{y_{0}}, \\ 
a_{2}:=\alpha _{5}x_{0}+i\alpha _{6}y_{0}+i\alpha _{7}p_{x_{0}}+\alpha
_{8}p_{y_{0}}, & a_{2}^{\dagger }:=\alpha _{5}x_{0}-i\alpha
_{6}y_{0}-i\alpha _{7}p_{x_{0}}+\alpha _{8}p_{y_{0}},%
\end{array}
\label{Fock}
\end{equation}%
with unknown constants $\alpha _{1},\ldots ,\alpha _{8}\in \mathbb{R}$ for
the time being, are $\mathcal{P}_{x}\mathcal{T}$-odd: $a_{i}\mapsto -a_{i}$, 
$a_{i}^{\dagger }\mapsto -a_{i}^{\dagger }$ and $\mathcal{P}_{y}\mathcal{T}$%
-even: $a_{i}\mapsto a_{i}$, $a_{i}^{\dagger }\mapsto a_{i}^{\dagger }$ when
using the realization (\ref{PT}). The reverse scenario is simply achieved by 
$\alpha _{j}\mapsto i\alpha _{j}$ for $j=1,\ldots ,8$.

The operators defined in (\ref{Fock}) satisfy the commutation relations (\ref%
{aaa}) provided that the following four constraints on the constants hold 
\begin{equation}
\alpha _{1}=\frac{\alpha _{6}}{2\hbar \Delta },\quad \ \alpha _{4}=\frac{%
\theta \alpha _{6}+\hbar \alpha _{7}}{2\hbar ^{2}\Delta },~~\quad \alpha
_{5}=-\frac{\alpha _{2}}{2\hbar \Delta },~~\quad \alpha _{8}=-\frac{\theta
\alpha _{2}+\hbar \alpha _{3}}{2\hbar ^{2}\Delta },  \label{ccc}
\end{equation}%
where we abbreviated $\Delta :=\alpha _{3}\alpha _{6}-\alpha _{2}\alpha
_{7}\neq 0$\footnote{%
For the specific choice 
\begin{equation*}
\alpha _{1}=\alpha _{2}=-\frac{\lambda _{1}}{\hbar \sqrt{K_{1}}},\qquad
\alpha _{3}=-\alpha _{4}=-\frac{1}{\sqrt{K_{1}}},\qquad \alpha _{5}=-\alpha
_{6}=\frac{\lambda _{2}}{\hbar \sqrt{K_{2}}},\qquad \alpha _{7}=\alpha _{8}=%
\frac{1}{\sqrt{K_{2}}},
\end{equation*}%
we recover the representation found in \cite{Scholtz:2008zu} when comparing
with equations (57) and (58) therein and identifying the quantities $\lambda
_{1},\lambda _{2}$ and $K_{1},K_{2}$ which are defined in equation (56) and
(59), respectively.}. This means we have still four almost entirely free
parameters left. Inverting the relations (\ref{Fock}) while keeping the
constraints (\ref{ccc}), we can express the coordinates and the momenta in
terms of the creation and annihilation operators 
\begin{equation}
\begin{array}{ll}
x_{0}=\left( \theta \alpha _{2}+\hbar \alpha _{3}\right)
(a_{1}+a_{1}^{\dagger })+\left( \theta \alpha _{6}+\hbar \alpha _{7}\right)
(a_{2}+a_{2}^{\dagger }),~ & y_{0}=\frac{i\alpha _{7}}{2\Delta }%
(a_{1}-a_{1}^{\dagger })-\frac{i\alpha _{3}}{2\Delta }(a_{2}-a_{2}^{\dagger
}), \\ 
p_{x_{0}}=-\frac{i\alpha _{6}}{2\Delta }(a_{1}-a_{1}^{\dagger })+\frac{%
i\alpha _{2}}{2\Delta }(a_{2}-a_{2}^{\dagger }), & p_{y_{0}}=-\hbar \alpha
_{2}(a_{1}+a_{1}^{\dagger })-\hbar \alpha _{6}(a_{2}+a_{2}^{\dagger }).%
\end{array}
\label{x0}
\end{equation}%
It is easily verified that these operators obey (\ref{1}) when using (\ref%
{aaa}).

\subsection{Oscillator algebras from string type noncommutative space-time}

Let us now carry out a similar analysis for the situation when the
underlying space-time is dynamical, i.e.~the constant $\theta $ becomes
position and possibly also momentum dependent. A set of consistent
commutation relations for such a scenario was introduced in \cite{AFLG} 
\begin{equation}
\begin{array}{lll}
\lbrack x,y]=i\theta (1+\tau y^{2}),\quad \quad & [x,p_{x}]=i\hbar (1+\tau
y^{2}), & [y,p_{y}]=i\hbar (1+\tau y^{2}), \\ 
\lbrack p_{x},p_{y}]=0, & [x,p_{y}]=2i\tau y(\theta p_{y}+\hbar x),\quad
\quad & [y,p_{x}]=0.%
\end{array}
\label{3}
\end{equation}
Defining the analogues to the creation and annihilation operators and
keeping the dependence on the dynamical variables similar as in (\ref{Fock}) 
\begin{equation}
\begin{array}{ll}
\hat{a}_{1}:=\alpha _{1}x+i\alpha _{2}y+i\alpha _{3}p_{x}+\alpha
_{4}p_{y},\qquad & \hat{a}_{1}^{\dagger }:=\alpha _{1}x-i\alpha
_{2}y-i\alpha _{3}p_{x}+\alpha _{4}p_{y}, \\ 
\hat{a}_{2}:=\alpha _{5}x+i\alpha _{6}y+i\alpha _{7}p_{x}+\alpha _{8}p_{y},
& \hat{a}_{2}^{\dagger }:=\alpha _{5}x-i\alpha _{6}y-i\alpha
_{7}p_{x}+\alpha _{8}p_{y},%
\end{array}
\label{28}
\end{equation}
we can compute the resulting commutation relations. Keeping the constraints (%
\ref{ccc}) and setting in addition $\alpha _{3}=0$ we find that the standard
commutation relations are deformed 
\begin{eqnarray}
\lbrack \hat{a}_{i},\hat{a}_{i}^{\dagger }] &=&1+\frac{\tau }{4\alpha
_{2}^{2}}\left( \hat{a}_{1}\hat{a}_{1}^{\dagger }+\hat{a}_{1}^{\dagger }\hat{%
a}_{1}-\hat{a}_{1}\hat{a}_{1}-\hat{a}_{1}^{\dagger }\hat{a}_{1}^{\dagger
}\right) \qquad \text{for }i=1,2  \label{def1} \\
\lbrack \hat{a}_{1},\hat{a}_{2}] &=&[\hat{a}_{1},\hat{a}_{2}^{\dagger }]=[%
\hat{a}_{1}^{\dagger },\hat{a}_{2}]=[\hat{a}_{1}^{\dagger },\hat{a}%
_{2}^{\dagger }]=\frac{\tau }{4\alpha _{2}^{2}}\left( \hat{a}_{1}\hat{a}_{2}+%
\hat{a}_{1}\hat{a}_{2}^{\dagger }-\hat{a}_{1}^{\dagger }\hat{a}_{2}-\hat{a}%
_{1}^{\dagger }\hat{a}_{2}^{\dagger }\right) .  \label{def2}
\end{eqnarray}
The asymmetry between $i=1$ and $i=2$ in (\ref{def1}) appears odd at first
sight in the light of (\ref{28}), but it is a consequence of the
non-symmetric nature of (\ref{3}) and our choice $\alpha _{3}=0$. Clearly
when the deformation parameter $\tau $ vanishes we obtain the usual Fock
space commutation relations (\ref{aaa}).

\subsection{Oscillator algebras from membrane type noncommutative space-time}

We propose now a new type of deformation for the flat noncommutative
space-time (\ref{1}) 
\begin{equation}
\!\!%
\begin{array}{lll}
\lbrack \tilde{x},\tilde{y}]=i\theta +i\tau \left( \tilde{x}^{2}+\tilde{y}%
^{2}\right) ,\quad \quad  & [\tilde{x},\tilde{p}_{x}]=i\hbar +i\frac{\tau
\hbar }{\theta }\left( \tilde{x}^{2}+\tilde{y}^{2}\right) , & [\tilde{x},%
\tilde{p}_{y}]=0, \\ 
\lbrack \tilde{p}_{x},\tilde{p}_{y}]=i\tau \left[ 2\frac{\hbar }{\theta }(%
\tilde{y}\tilde{p}_{x}-\tilde{x}\tilde{p}_{y})-\tilde{p}_{x}^{2}-\tilde{p}%
_{y}^{2}\right] ,~ & [\tilde{y},\tilde{p}_{y}]=i\hbar +i\frac{\tau \hbar }{%
\theta }\left( \tilde{x}^{2}+\tilde{y}^{2}\right) ,~ & [\tilde{y},\tilde{p}%
_{x}]=0.%
\end{array}
\label{tau}
\end{equation}%
In the same manner as for (\ref{3}) we may verify that these commutation
relations are consistent in the sense that the Jacobi identities are
satisfied. Using the standard arguments to find a minimal length, we observe
that the $\tilde{x},\tilde{y}$-commutator implies a minimal length in the $%
\tilde{x}$ as well as in the $\tilde{y}$-direction, which means the
underlying object, whose substructure we can not determine, is of a membrane
structure. Once again we define creation and annihilation type operators
analogously to (\ref{Fock}) keeping the dependence on the dynamical
variables the same. When specifying the coefficients such that 
\begin{equation}
\begin{array}{ll}
\tilde{a}_{1}:=\sqrt{\frac{1-\tau }{2\theta }}(\tilde{x}+i\tilde{y}),\qquad 
& \tilde{a}_{1}^{\dagger }:=\sqrt{\frac{1-\tau }{2\theta }}(\tilde{x}-i%
\tilde{y}), \\ 
\tilde{a}_{2}:=\sqrt{\frac{1-\tau }{2\theta }}\left[ \tilde{x}-i\tilde{y}+%
\frac{\theta }{\hbar }(\tilde{p}_{y}+i\tilde{p}_{x})\right] ,~~~~~~ & \tilde{%
a}_{2}^{\dagger }:=\sqrt{\frac{1-\tau }{2\theta }}\left[ \tilde{x}+i\tilde{y}%
+\frac{\theta }{\hbar }(\tilde{p}_{y}-i\tilde{p}_{x})\right] ,%
\end{array}%
\end{equation}%
we find the commutation relations 
\begin{equation}
\tilde{a}_{i}\tilde{a}_{j}^{\dagger }-\left( \frac{1+\tau }{1-\tau }\right)
^{\delta _{ij}}\tilde{a}_{j}^{\dagger }\tilde{a}_{i}=\delta _{ij},\quad
\lbrack \tilde{a}_{i}^{\dagger },\tilde{a}_{j}^{\dagger }]=0,\quad \lbrack 
\tilde{a}_{i},\tilde{a}_{j}]=0,\qquad \text{for }i,j=1,2.  \label{ast}
\end{equation}%
As expected (\ref{aaa}) is recovered for $\tau \rightarrow 0$. These
relations are very reminiscent of the q-deformed oscillator algebra studied
in this context for instance in \cite%
{Kempf1,Kempf2,Brodimas,Bieden,MacF,ChangPu,QuesneTK,AFBB}.

This example and the one in the previous subsection indicate that dynamical
space-time relations will naturally lead to deformed Fock spaces. As we have
seen some of them have a very convenient and well studied structure, as (\ref%
{ast}), whereas others are rather awkward such as (\ref{def1}) and (\ref%
{def2}). Let us therefore now reverse the scenario and deform first the Fock
space relations in a \textquotedblleft nice\textquotedblright\ way and
subsequently compute the corresponding commutation relations for the
dynamical variables.

\section{Noncommutative space-time from q-deformed creation and annihilation
operators}

Resembling the relations (\ref{ast}) we $q$-deform the relations in (\ref%
{aaa}) by defining a new set of creation and annihilation operators $%
A_{1},A_{1}^{\dagger },A_{2},A_{2}^{\dagger }$ satisfying 
\begin{equation}
A_{i}A_{j}^{\dagger }-q^{2\delta _{ij}}A_{j}^{\dagger }A_{i}=\delta
_{ij},\quad \lbrack A_{i}^{\dagger },A_{j}^{\dagger }]=0,\quad \lbrack
A_{i},A_{j}]=0,\qquad \text{for }i,j=1,2.  \label{AAAA}
\end{equation}%
There exist various other possibilities to deform the relations (\ref{aaa})
which still lead to constructable Fock spaces, such as for instance using
different $q$s in the first relation of (\ref{AAAA}), i.e.~$q^{2\delta
_{ij}}\rightarrow q_{i}^{2\delta _{ij}}$ or replacing the $\delta _{ij}$ on
the right hand side of the first relation by $q^{g(A_{i}^{\dagger }A_{i})}$
with $g(x)$ being an arbitrary function as in \cite{Brodimas,AFBB}. Guided
by the limit $q\rightarrow 1$ in which we should recover the relations (\ref%
{x0}) and the properties of these operators under a $\mathcal{PT}$%
-transformation, we expand the new set of deformed canonical variables $%
X,Y,P_{x},P_{y}$ linearly in terms of the $A_{1},A_{1}^{\dagger
},A_{2},A_{2}^{\dagger }$ as 
\begin{equation}
\begin{array}{ll}
X=\kappa _{1}(A_{1}^{\dagger }+A_{1})+\kappa _{2}(A_{2}^{\dagger
}+A_{2}),\qquad  & P_{x}=i\kappa _{3}(A_{1}^{\dagger }-A_{1})+i\kappa
_{4}(A_{2}^{\dagger }-A_{2}), \\ 
Y=i\kappa _{5}(A_{1}^{\dagger }-A_{1})+i\kappa _{6}(A_{2}^{\dagger }-A_{2}),
& P_{y}=\kappa _{7}(A_{1}^{\dagger }+A_{1})+\kappa _{8}(A_{2}^{\dagger
}+A_{2}).%
\end{array}
\label{XP}
\end{equation}%
The constants $\kappa _{1},\ldots ,\kappa _{8}\in \mathbb{R}$ are unknown
for the time being. Inverting the relations (\ref{XP}) we may express the
deformed creation and annihilation operators in terms of the deformed
canonical variables 
\begin{equation}
\begin{array}{ll}
A_{1}=\frac{\kappa _{8}}{\lambda }X+i\frac{\kappa _{4}}{\mu }Y-i\frac{\kappa
_{6}}{\mu }P_{x}-\frac{\kappa _{2}}{\lambda }P_{y},\qquad  & A_{1}^{\dagger
}=\frac{\kappa _{8}}{\lambda }X-i\frac{\kappa _{4}}{\mu }Y+i\frac{\kappa _{6}%
}{\mu }P_{x}-\frac{\kappa _{2}}{\lambda }P_{y}, \\ 
A_{2}=-\frac{\kappa _{7}}{\lambda }X-i\frac{\kappa _{3}}{\mu }Y+i\frac{%
\kappa _{5}}{\mu }P_{x}+\frac{\kappa _{1}}{\lambda }P_{y}, & A_{2}^{\dagger
}=-\frac{\kappa _{7}}{\lambda }X+i\frac{\kappa _{3}}{\mu }Y-i\frac{\kappa
_{5}}{\mu }P_{x}+\frac{\kappa _{1}}{\lambda }P_{y},%
\end{array}
\label{AA}
\end{equation}%
where we abbreviated $\lambda :=2(\kappa _{1}\kappa _{8}-\kappa _{2}\kappa
_{7})\neq 0$ and $\mu :=2(\kappa _{4}\kappa _{5}-\kappa _{3}\kappa _{6})\neq
0$. Using the representation (\ref{XP}) together with (\ref{AAAA}) we
compute 
\begin{eqnarray}
\lbrack X,Y] &=&2i(\kappa _{1}\kappa _{5}+\kappa _{2}\kappa
_{6})+2i(q^{2}-1)(\kappa _{1}\kappa _{5}A_{1}^{\dagger }A_{1}+\kappa
_{2}\kappa _{6}A_{2}^{\dagger }A_{2}),  \label{c1} \\
\lbrack X,P_{x}] &=&2i(\kappa _{1}\kappa _{3}+\kappa _{2}\kappa
_{4})+2i(q^{2}-1)(\kappa _{1}\kappa _{3}A_{1}^{\dagger }A_{1}+\kappa
_{2}\kappa _{4}A_{2}^{\dagger }A_{2}), \\
\lbrack Y,P_{y}] &=&-2i(\kappa _{5}\kappa _{7}+\kappa _{6}\kappa
_{8})+2i(1-q^{2})(\kappa _{5}\kappa _{7}A_{1}^{\dagger }A_{1}+\kappa
_{6}\kappa _{8}A_{2}^{\dagger }A_{2}), \\
\lbrack P_{x},P_{y}] &=&-2i(\kappa _{3}\kappa _{7}+\kappa _{4}\kappa
_{8})+2i(1-q^{2})(\kappa _{3}\kappa _{7}A_{1}^{\dagger }A_{1}+\kappa
_{4}\kappa _{8}A_{2}^{\dagger }A_{2}),  \label{c4} \\
\lbrack X,P_{y}] &=&0,  \label{c5} \\
\lbrack Y,P_{x}] &=&0.  \label{c6}
\end{eqnarray}%
Next we employ the relations (\ref{AA}) and evaluate 
\begin{eqnarray}
A_{1}^{\dagger }A_{1} &=&\frac{\kappa _{8}^{2}}{\lambda ^{2}}X^{2}+\frac{%
\kappa _{4}^{2}}{\mu ^{2}}Y^{2}+\frac{\kappa _{6}^{2}}{\mu ^{2}}P_{x}^{2}+%
\frac{\kappa _{2}^{2}}{\lambda ^{2}}P_{y}^{2}-\frac{2\kappa _{8}\kappa _{2}}{%
\lambda ^{2}}XP_{y}-\frac{2\kappa _{4}\kappa _{6}}{\mu ^{2}}YP_{x}
\label{aa1} \\
&&+i\frac{\kappa _{4}\kappa _{8}}{\lambda \mu }[X,Y]+i\frac{\kappa
_{4}\kappa _{2}}{\lambda \mu }[Y,P_{y}]-i\frac{\kappa _{6}\kappa _{8}}{%
\lambda \mu }[X,P_{x}]-i\frac{\kappa _{6}\kappa _{2}}{\lambda \mu }%
[P_{x},P_{y}],  \notag \\
A_{2}^{\dagger }A_{2} &=&\frac{\kappa _{7}^{2}}{\lambda ^{2}}X^{2}+\frac{%
\kappa _{3}^{2}}{\mu ^{2}}Y^{2}+\frac{\kappa _{5}^{2}}{\mu ^{2}}P_{x}^{2}+%
\frac{\kappa _{1}^{2}}{\lambda ^{2}}P_{y}^{2}-\frac{2\kappa _{7}\kappa _{1}}{%
\lambda ^{2}}XP_{y}-\frac{2\kappa _{3}\kappa _{5}}{\mu ^{2}}YP_{x}
\label{aa2} \\
&&+i\frac{\kappa _{3}\kappa _{7}}{\lambda \mu }[X,Y]+i\frac{\kappa
_{3}\kappa _{1}}{\lambda \mu }[Y,P_{y}]-i\frac{\kappa _{5}\kappa _{7}}{%
\lambda \mu }[X,P_{x}]-i\frac{\kappa _{5}\kappa _{1}}{\lambda \mu }%
[P_{x},P_{y}].  \notag
\end{eqnarray}%
Substituting (\ref{aa1}) and (\ref{aa2}) into the right hand sides of (\ref%
{c1})-(\ref{c4}) we obtain four equations for the four unknown commutators $%
[X,Y]$, $[X,P_{x}]$, $[Y,P_{y}]$ and $[P_{x},P_{y}]$. Solving these
equations, the resulting dynamical noncommutative relations are 
\begin{eqnarray}
\lbrack X,Y] &=&i\theta +i\frac{q-q^{-1}}{q+q^{-1}}\left[ \frac{\kappa
_{2}\kappa _{6}\kappa _{7}^{2}+\kappa _{1}\kappa _{5}\kappa _{8}^{2}}{\left(
\kappa _{2}\kappa _{7}-\kappa _{1}\kappa _{8}\right) ^{2}}X^{2}+\frac{\kappa
_{2}\kappa _{6}\kappa _{3}^{2}+\kappa _{1}\kappa _{5}\kappa _{4}^{2}}{\left(
\kappa _{4}\kappa _{5}-\kappa _{3}\kappa _{6}\right) ^{2}}Y^{2}\right. ~~~
\label{XY} \\
&&+\frac{\kappa _{5}\kappa _{6}\left( \kappa _{2}\kappa _{5}+\kappa
_{1}\kappa _{6}\right) }{\left( \kappa _{4}\kappa _{5}-\kappa _{3}\kappa
_{6}\right) ^{2}}P_{x}^{2}+\frac{\kappa _{1}\kappa _{2}\left( \kappa
_{2}\kappa _{5}+\kappa _{1}\kappa _{6}\right) }{\left( \kappa _{2}\kappa
_{7}-\kappa _{1}\kappa _{8}\right) ^{2}}P_{y}^{2}  \notag \\
&&-\left. \frac{2\kappa _{1}\kappa _{2}\left( \kappa _{6}\kappa _{7}+\kappa
_{5}\kappa _{8}\right) }{\left( \kappa _{2}\kappa _{7}-\kappa _{1}\kappa
_{8}\right) ^{2}}XP_{y}-\frac{2\kappa _{5}\kappa _{6}\left( \kappa
_{2}\kappa _{3}+\kappa _{1}\kappa _{4}\right) }{\left( \kappa _{4}\kappa
_{5}-\kappa _{3}\kappa _{6}\right) ^{2}}YP_{x}\right] ,  \notag \\
\lbrack X,P_{x}] &=&i\hbar +i\frac{q-q^{-1}}{q+q^{-1}}\left[ \frac{\kappa
_{2}\kappa _{4}\kappa _{7}^{2}+\kappa _{1}\kappa _{3}\kappa _{8}^{2}}{\left(
\kappa _{2}\kappa _{7}-\kappa _{1}\kappa _{8}\right) ^{2}}X^{2}+\frac{\kappa
_{3}\kappa _{4}\left( \kappa _{2}\kappa _{3}+\kappa _{1}\kappa _{4}\right) }{%
\left( \kappa _{4}\kappa _{5}-\kappa _{3}\kappa _{6}\right) ^{2}}%
Y^{2}\right. ~~~ \\
&&+\frac{\kappa _{2}\kappa _{4}\kappa _{5}^{2}+\kappa _{1}\kappa _{3}\kappa
_{6}^{2}}{\left( \kappa _{4}\kappa _{5}-\kappa _{3}\kappa _{6}\right) ^{2}}%
P_{x}^{2}+\frac{\kappa _{1}\kappa _{2}\left( \kappa _{2}\kappa _{3}+\kappa
_{1}\kappa _{4}\right) }{\left( \kappa _{2}\kappa _{7}-\kappa _{1}\kappa
_{8}\right) ^{2}}P_{y}^{2}  \notag \\
&&-\left. \frac{2\kappa _{1}\kappa _{2}\left( \kappa _{4}\kappa _{7}+\kappa
_{3}\kappa _{8}\right) }{\left( \kappa _{2}\kappa _{7}-\kappa _{1}\kappa
_{8}\right) ^{2}}XP_{y}-\frac{2\kappa _{3}\kappa _{4}\left( \kappa
_{2}\kappa _{5}+\kappa _{1}\kappa _{6}\right) }{\left( \kappa _{4}\kappa
_{5}-\kappa _{3}\kappa _{6}\right) ^{2}}YP_{x}\right] ,  \notag \\
\lbrack Y,P_{y}] &=&i\hbar -i\frac{q-q^{-1}}{q+q^{-1}}\left[ \frac{\kappa
_{7}\kappa _{8}\left( \kappa _{6}\kappa _{7}+\kappa _{5}\kappa _{8}\right) }{%
\left( \kappa _{2}\kappa _{7}-\kappa _{1}\kappa _{8}\right) ^{2}}X^{2}+\frac{%
\kappa _{6}\kappa _{8}\kappa _{3}^{2}+\kappa _{5}\kappa _{7}\kappa _{4}^{2}}{%
\left( \kappa _{4}\kappa _{5}-\kappa _{3}\kappa _{6}\right) ^{2}}%
Y^{2}\right. ~~~ \\
&&+\frac{\kappa _{5}\kappa _{6}\left( \kappa _{6}\kappa _{7}+\kappa
_{5}\kappa _{8}\right) }{\left( \kappa _{4}\kappa _{5}-\kappa _{3}\kappa
_{6}\right) ^{2}}P_{x}^{2}+\frac{\kappa _{6}\kappa _{8}\kappa
_{1}^{2}+\kappa _{5}\kappa _{7}\kappa _{2}^{2}}{\left( \kappa _{2}\kappa
_{7}-\kappa _{1}\kappa _{8}\right) ^{2}}P_{y}^{2}  \notag \\
&&-\left. \frac{2\kappa _{7}\kappa _{8}\left( \kappa _{2}\kappa _{5}+\kappa
_{1}\kappa _{6}\right) }{\left( \kappa _{2}\kappa _{7}-\kappa _{1}\kappa
_{8}\right) ^{2}}XP_{y}-\frac{2\kappa _{5}\kappa _{6}\left( \kappa
_{4}\kappa _{7}+\kappa _{3}\kappa _{8}\right) }{\left( \kappa _{4}\kappa
_{5}-\kappa _{3}\kappa _{6}\right) ^{2}}YP_{x}\right] ,  \notag \\
\lbrack P_{x},P_{y}] &=&-i\frac{q-q^{-1}}{q+q^{-1}}\left[ \frac{\kappa
_{7}\kappa _{8}\left( \kappa _{4}\kappa _{7}+\kappa _{3}\kappa _{8}\right) }{%
\left( \kappa _{2}\kappa _{7}-\kappa _{1}\kappa _{8}\right) ^{2}}X^{2}+\frac{%
\kappa _{3}\kappa _{4}\left( \kappa _{4}\kappa _{7}+\kappa _{3}\kappa
_{8}\right) }{\left( \kappa _{4}\kappa _{5}-\kappa _{3}\kappa _{6}\right)
^{2}}Y^{2}\right. ~~~  \label{pxpy} \\
&&+\frac{\kappa _{4}\kappa _{8}\kappa _{5}^{2}+\kappa _{3}\kappa _{7}\kappa
_{6}^{2}}{\left( \kappa _{4}\kappa _{5}-\kappa _{3}\kappa _{6}\right) ^{2}}%
P_{x}^{2}+\frac{\kappa _{4}\kappa _{8}\kappa _{1}^{2}+\kappa _{3}\kappa
_{7}\kappa _{2}^{2}}{\left( \kappa _{2}\kappa _{7}-\kappa _{1}\kappa
_{8}\right) ^{2}}P_{y}^{2}  \notag \\
&&-\left. \frac{2\kappa _{7}\kappa _{8}\left( \kappa _{2}\kappa _{3}+\kappa
_{1}\kappa _{4}\right) }{\left( \kappa _{2}\kappa _{7}-\kappa _{1}\kappa
_{8}\right) ^{2}}XP_{y}-\frac{2\kappa _{3}\kappa _{4}\left( \kappa
_{6}\kappa _{7}+\kappa _{5}\kappa _{8}\right) }{\left( \kappa _{4}\kappa
_{5}-\kappa _{3}\kappa _{6}\right) ^{2}}YP_{x}\right] .  \notag
\end{eqnarray}%
For the constant terms of these commutators we have implemented here the
constraints 
\begin{eqnarray}
\kappa _{1}\kappa _{5}+\kappa _{2}\kappa _{6} &=&\frac{\theta }{4}\left(
1+q^{2}\right) ,  \label{kk1} \\
\kappa _{1}\kappa _{3}+\kappa _{2}\kappa _{4} &=&\frac{\hbar }{4}\left(
1+q^{2}\right) ,  \label{kk2} \\
\kappa _{5}\kappa _{7}+\kappa _{6}\kappa _{8} &=&-\frac{\hbar }{4}\left(
1+q^{2}\right) ,  \label{kk3} \\
\kappa _{3}\kappa _{7}+\kappa _{4}\kappa _{8} &=&0,  \label{kk}
\end{eqnarray}%
in order to ensure that the limit $q\rightarrow 1$ for the relations (\ref%
{XY})-(\ref{pxpy}) will yield the standard commutation relations for
noncommutative flat space-time (\ref{1}). The relations (\ref{c5}) and (\ref%
{c6}) remain of course unchanged.

\subsection{Some special limits}

Keeping all the constants generic in the algebra (\ref{XY})-(\ref{pxpy})
will make the handling very cumbersome. However, using the fact that we
still have four $\kappa $s free at our disposal allows us to extract some
special limiting cases in order to obtain some more tractable algebras.

\subsubsection{Dependent X and Y directions}

Considering (\ref{XP}) the first natural limit is to reduce the number of
free parameters to four, e.g.~$\kappa _{1},\ldots ,\kappa _{4}$, and
introduce some dependence for the coefficients in the $Y$-direction on those
in the $X$-direction. Considering the representation (\ref{AA}) we impose 
\begin{equation}
\kappa _{5}=\kappa _{1},\qquad \kappa _{6}=-\kappa _{2},\qquad \kappa
_{7}=-\kappa _{3}\qquad \text{and\qquad }\kappa _{8}=\kappa _{4},  \label{cv}
\end{equation}%
such that without activating the constraints (\ref{kk1})-(\ref{kk}) the
eight unknown constants are already limited to four. The four constraints (%
\ref{kk1})-(\ref{kk}) are not independent for these choices as (\ref{kk2})
and (\ref{kk3}) become identical. The remaining three constraints read 
\begin{equation}
\kappa _{1}^{2}-\kappa _{2}^{2}=\frac{\theta }{4}\left( 1+q^{2}\right)
,\qquad \kappa _{1}\kappa _{3}+\kappa _{2}\kappa _{4}=\frac{\hbar }{4}\left(
1+q^{2}\right) \quad \text{and\quad }\kappa _{3}^{2}=\kappa _{4}^{2},
\label{hh3}
\end{equation}%
which means we have still one constant at our disposal. The algebra (\ref{XY}%
)-(\ref{pxpy}), (\ref{c5}) and (\ref{c6}) simplifies to 
\begin{eqnarray}
\lbrack X,Y] &=&i\theta +i\frac{q-q^{-1}}{q+q^{-1}}\left[ \frac{\kappa
_{1}\kappa _{4}-\kappa _{2}\kappa _{3}}{\kappa _{1}\kappa _{4}+\kappa
_{2}\kappa _{3}}(X^{2}+Y^{2})-\frac{2\kappa _{1}\kappa _{2}}{\kappa
_{1}\kappa _{4}+\kappa _{2}\kappa _{3}}(XP_{y}-YP_{x})\right] ,~\ \ ~
\label{v1} \\
\lbrack X,P_{x}] &=&ih+i\frac{q-q^{-1}}{q+q^{-1}}\left[ \frac{\kappa
_{3}\kappa _{4}}{\kappa _{1}\kappa _{4}+\kappa _{2}\kappa _{3}}(X^{2}+Y^{2})+%
\frac{\kappa _{1}\kappa _{2}}{\kappa _{1}\kappa _{4}+\kappa _{2}\kappa _{3}}%
(P_{x}^{2}+P_{y}^{2})\right] , \\
\lbrack Y,P_{y}] &=&ih+i\frac{q-q^{-1}}{q+q^{-1}}\left[ \frac{\kappa
_{3}\kappa _{4}}{\kappa _{1}\kappa _{4}+\kappa _{2}\kappa _{3}}(X^{2}+Y^{2})+%
\frac{\kappa _{1}\kappa _{2}}{\kappa _{1}\kappa _{4}+\kappa _{2}\kappa _{3}}%
(P_{x}^{2}+P_{y}^{2})\right] , \\
\lbrack P_{x},P_{y}] &=&-i\frac{q-q^{-1}}{q+q^{-1}}\left[ \frac{\kappa
_{1}\kappa _{4}-\kappa _{2}\kappa _{3}}{\kappa _{1}\kappa _{4}+\kappa
_{2}\kappa _{3}}(P_{x}^{2}+P_{y}^{2})-\frac{2\kappa _{3}\kappa _{4}}{\kappa
_{1}\kappa _{4}+\kappa _{2}\kappa _{3}}(XP_{y}-YP_{x})\right] , \\
\lbrack X,P_{y}] &=&0, \\
\lbrack Y,P_{x}] &=&0.  \label{v6}
\end{eqnarray}%
The conditions $\lambda \neq 0$, $\mu \neq 0$ now coincide and have
translated into $\kappa _{1}\kappa _{4}+\kappa _{2}\kappa _{3}\neq 0$. Our
choice of constants has achieved that the terms $XP_{y}$ and $YP_{x}$ have
combined into the angular momentum operator $L_{z}$.

\subsection{Membrane and string type relations}

As one of the $\kappa $s is still not fixed we can simplify the commutation
relations (\ref{v1})-(\ref{v6}) further by setting $\kappa _{2}=0$, such
that all three unknown left are fixed by the remaining three relations 
\begin{equation}
\kappa _{1}^{2}=\frac{\theta }{4}\left( 1+q^{2}\right) ,\qquad \kappa
_{1}\kappa _{3}=\frac{\hbar }{4}\left( 1+q^{2}\right) \quad \text{and\quad }%
\kappa _{3}^{2}=\kappa _{4}^{2}.  \label{kappa}
\end{equation}%
We may now implement the constraints (\ref{kappa}) in the algebra (\ref{v1}%
)-(\ref{v6}) and eliminate all constants $\kappa _{i}$ being left with a
purely $q$-deformed algebra 
\begin{eqnarray}
\lbrack X,Y] &=&i\theta +i\frac{q-q^{-1}}{q+q^{-1}}\left( X^{2}+Y^{2}\right)
,~\ \ ~  \label{n1} \\
\lbrack X,P_{x}] &=&i\hbar +i\frac{q-q^{-1}}{q+q^{-1}}\frac{\hbar }{\theta }%
\left( X^{2}+Y^{2}\right) , \\
\lbrack Y,P_{y}] &=&i\hbar +i\frac{q-q^{-1}}{q+q^{-1}}\frac{\hbar }{\theta }%
\left( X^{2}+Y^{2}\right) , \\
\lbrack P_{x},P_{y}] &=&i\frac{q^{-1}-q}{q^{-1}+q}\left[
P_{x}^{2}+P_{y}^{2}+2\frac{\hbar }{\theta }(XP_{y}-YP_{x})\right] , \\
\lbrack X,P_{y}] &=&0, \\
\lbrack Y,P_{x}] &=&0.  \label{n6}
\end{eqnarray}%
These relations reduce to (\ref{tau}) for $q=\pm \sqrt{(1+\tau )/(1-\tau )}$%
. Notice further that the $q$-deformation and the $\theta $-deformation
originally introduced in the space-space commutation relations have become
intrinsically linked through the constraints. We can no longer take the
limit $\theta \rightarrow 0$ separately without taking also the limit $%
q\rightarrow 0$. However, the limit $q\rightarrow 0$ may still be taken
separately and we recover (\ref{1}).

We named these relations ``membrane type''\ as the relation (\ref{n1}) will
give rise to a minimal length in the $X$ and $Y$ direction in a simultaneous
measurement as we will explain in more detail below. As it stands, the
relation (\ref{n1}) will lead to the same minimal length in either
direction. This is by no means unavoidable and can be overcome by taking
another limit of the algebra (\ref{XY})-(\ref{pxpy}), (\ref{c5}) and (\ref%
{c6}). Setting for instance $\kappa _{2}=\kappa _{6}=0$ without any
additional constraints besides (\ref{kk1})-(\ref{kk}), which in this case
read 
\begin{equation}
\kappa _{1}\kappa _{5}=\frac{\theta }{4}\left( 1+q^{2}\right) ,\quad \kappa
_{1}\kappa _{3}=\frac{\hbar }{4}\left( 1+q^{2}\right) ,\quad \kappa
_{5}\kappa _{7}=-\frac{\hbar }{4}\left( 1+q^{2}\right) ,\quad \kappa
_{3}\kappa _{7}=-\kappa _{4}\kappa _{8}.  \label{ca}
\end{equation}
the algebra simplifies considerably 
\begin{eqnarray}
\lbrack X,Y] &=&i\theta +i\frac{q-q^{-1}}{q+q^{-1}}\left( \frac{\kappa _{5}}{%
\kappa _{1}}X^{2}+\frac{\kappa _{1}}{\kappa _{5}}Y^{2}\right) ,  \label{11}
\\
\lbrack X,P_{x}] &=&i\hbar +i\frac{q-q^{-1}}{q+q^{-1}}\left( \frac{\kappa
_{3}}{\kappa _{1}}X^{2}+\frac{\kappa _{1}\kappa _{3}}{\kappa _{5}^{2}}%
Y^{2}\right) , \\
\lbrack Y,P_{y}] &=&i\hbar -i\frac{q-q^{-1}}{q+q^{-1}}\left( \frac{\kappa
_{5}\kappa _{7}}{\kappa _{1}^{2}}X^{2}+\frac{\kappa _{7}}{\kappa _{5}}%
Y^{2}\right) , \\
\lbrack P_{x},P_{y}] &=&-i\frac{q-q^{-1}}{q+q^{-1}}\left[ (\kappa _{4}\kappa
_{7}+\kappa _{3}\kappa _{8})\left( \frac{\kappa _{7}}{\kappa _{8}\kappa
_{1}^{2}}X^{2}+\frac{\kappa _{3}}{\kappa _{4}\kappa _{5}^{2}}Y^{2}\right)
\right.  \label{as} \\
&&\left. +\frac{\kappa _{8}}{\kappa _{4}}P_{x}^{2}+\frac{\kappa _{4}}{\kappa
_{8}}P_{y}^{2}-2\frac{\kappa _{4}\kappa _{7}}{\kappa _{1}\kappa _{8}}YP_{x}-2%
\frac{\kappa _{3}\kappa _{8}}{\kappa _{4}\kappa _{5}}XP_{y}\right] ,  \notag
\\
\lbrack X,P_{y}] &=&0, \\
\lbrack Y,P_{x}] &=&0.  \label{6}
\end{eqnarray}
We notice that in (\ref{11}) we have now different coefficients in front of
the $X^{2}$ and $Y^{2}$-terms and may achieve unequal minimal length in
either direction, although they are not entirely independent being related
by the first relation in (\ref{ca}).

Taking now a less trivial limit, we may obtain string like relations from (%
\ref{11})-(\ref{6}) similar to those proposed in \cite{AFLG}. Parameterizing 
$q=e^{2\tau \kappa _{5}^{2}}$ with $\tau \in \mathbb{R}^{+}$ and taking the
limit $\kappa _{5}\rightarrow 0$ we obtain yet simpler relations. As we have
still many free parameters left in (\ref{as}) we have several choices. With
respect to the constraints (\ref{ca}) we can take for instance $\kappa
_{3}=\hbar /\theta \kappa _{5}$, $\kappa _{4}=\hbar ^{2}/\theta \kappa _{5}$%
, $\kappa _{8}=(1+q^{2})/(4\kappa _{5})$ and derive the simple
\textquotedblleft string type\textquotedblright\ relations 
\begin{equation}
\begin{array}{lll}
\lbrack X,Y]=i\theta \left( 1+\tau Y^{2}\right) ,\qquad & [X,P_{x}]=i\hbar
\left( 1+\tau Y^{2}\right) ,\qquad & [X,P_{y}]=0, \\ 
\lbrack P_{x},P_{y}]=i\tau \frac{\hbar ^{2}}{\theta }Y^{2},~~ & 
[Y,P_{y}]=i\hbar \left( 1+\tau Y^{2}\right) ,~\  & [Y,P_{x}]=0.%
\end{array}
\label{bbb}
\end{equation}%
Arguing in the same way as in \cite{AFLG}, we obtain now from the first
relation in (\ref{bbb}) a minimal length in the $Y$-direction in a
simultaneous $X,Y$-measurement as the\ commutator $[X,Y]$ is identical. The
remaining commutators are, however, different.

There are of course plenty of other possible limits compatible with the
constraints (\ref{kk1})-(\ref{kk}), which we do not present here.

\section{Minimal areas and minimal lengths}

As mentioned, one of the interesting physical consequences of noncommutative
space-time, especially when it is dynamical, is the emergence of minimal
lengths in simultaneous measurements of two observables. The standard
noncommutative space-time relations (\ref{1}) give rise to additional
uncertainties similar to the usual Heisenberg uncertainty relations, meaning
for instance that the two position operators $x_{0}$ and $y_{0}$ can never
be known with complete precision at \emph{the same time}, where $\theta $
plays the role of $\hbar $ when compared with the conventional relations.
When the underlying algebra becomes a dynamical noncommutative space-time
structure the consequences are more severe and one finds that the position
operators $X$ or $Y$ can \emph{never} be known, that is even when giving up
the entire knowledge about the canonical conjugate partner $Y$ or $X$,
respectively. Thus $X$ or $Y$ are said to be bound by some absolute minimal
length $\Delta X_{0}$ or $\Delta Y_{0}$, which is the highest possible
precision to which these quantities can be resolved.

Minimal lengths have been known and studied for some time \cite%
{Kempf1,Kempf2,Brodimas,Bieden,MacF,ChangPu,QuesneTK,AFBB} in simultaneous $%
x,p$-measurements as a consequence of a deformation of the $x,p$-commutator.
In \cite{AFLG} it was demonstrated explicitly that they also result in
simultaneous $x,y$-measurements as a consequence of the dynamical
noncommutativity of space-time. Whereas the algebra investigated in \cite%
{AFLG} only gave rise to a minimal length in one direction, i.e.~``string
like''\ objects, we demonstrate here that the algebras provided in section 3
will lead to minimal lengths in two direction, i.e.~minimal areas. Objects
in these type of spaces are ``membrane like'', meaning that there exists a
finitely extended region about whose substructure it is impossible to obtain
any measurable knowledge.

Following the standard arguments we will now compute these quantities by
starting with the well known relation 
\begin{equation}
\Delta A\Delta B\geq \frac{1}{2}\left\vert \left\langle [A,B]\right\rangle
\right\vert ,  \label{HU}
\end{equation}%
which holds for any two observables $A$ and $B$, which are Hermitian with
respect to the standard inner product. In order to determine the range of
validity for this inequality we simply have to minimize $f(\Delta A,\Delta
B):=\Delta A\Delta B-\frac{1}{2}\left\vert \left\langle [A,B]\right\rangle
\right\vert $ as a function of $\Delta B$ to find the absolute minimal
length $\Delta A_{0}$. This means we need to solve the two equations $%
\partial _{\Delta B}f(\Delta A,\Delta B)=0$ and $f(\Delta A,\Delta B)=0$ for 
$\Delta A=:$ $\Delta A_{\text{min}}$ and subsequently compute the smallest
value for $\Delta A_{\text{min}}$ in order to obtain the absolute minimal
length $\Delta A_{0}$. In case we obtain minimal length for both of these
observables we define the minimal area and its smallest possible value of
four times the product, that is $\Delta (AB)_{\text{min}}$ and $\Delta
(AB)_{0}$, respectively.

For definiteness we choose now $\theta \in \mathbb{R}^{+}$ and carry out the
analysis for the algebra (\ref{11})-(\ref{6}) starting with a simultaneous $%
X,Y$-measurement. When $q^{2}>1$ the imaginary parts of all terms of the
commutator $[X,Y]$ are positive due to the first constraint in (\ref{ca}).
The absolute value for $\left\vert \left\langle [X,Y]\right\rangle
\right\vert $ is therefore simply $\func{Im}\left\langle [X,Y]\right\rangle $%
. When $q^{2}<1$ we use $\left\vert A-B\right\vert \geq A-B$ for $A,B>0$ to
drop the absolute value. Using furthermore that the mean-squared deviation
about the expectation value $\left\langle X\right\rangle $ is given by $%
\Delta X^{2}=\left\langle X^{2}\right\rangle -\left\langle X\right\rangle
^{2}$ and similarly for $X\leftrightarrow Y$, we compute 
\begin{eqnarray}
\Delta X_{\text{min}} &=&\frac{\sqrt{\left\vert q^{2}-1\right\vert (\kappa
_{1}^{2}\left\langle X\right\rangle ^{2}+\kappa _{5}^{2}\left\langle
Y\right\rangle ^{2})+\theta (q^{4}-1)\kappa _{1}\kappa _{5}}}{2q\kappa _{5}},
\\
\Delta Y_{\text{min}} &=&\frac{\sqrt{\left\vert q^{2}-1\right\vert (\kappa
_{5}^{2}\left\langle X\right\rangle ^{2}+\kappa _{1}^{2}\left\langle
Y\right\rangle ^{2})+\theta (q^{4}-1)\kappa _{1}\kappa _{5}}}{2q\kappa _{1}},
\end{eqnarray}%
such that the absolute minimal lengths result to 
\begin{equation}
\Delta X_{0}=\frac{\kappa _{1}}{q}\sqrt{\left\vert q^{2}-1\right\vert }%
\qquad \text{and\qquad }\Delta Y_{0}=\frac{\kappa _{5}}{q}\sqrt{\left\vert
q^{2}-1\right\vert },  \label{XY0}
\end{equation}%
hen $\left\langle X\right\rangle =\left\langle Y\right\rangle =0$. Together
with the first constraint in (\ref{ca}) the absolute minimal area in the $%
X,Y $-plane results to 
\begin{equation}
\Delta (XY)_{0}=\theta \left\vert q^{2}-q^{-2}\right\vert .  \label{mina}
\end{equation}%
This means the size of the minimal area is independent of the free
parameters $\kappa _{1}$ and $\kappa _{5}$. We can also make $\Delta Y_{0}$
a function of $\Delta X_{0}$ and compute for given $\Delta X_{0}$ the
corresponding minimal length $\Delta Y_{0}$ or vice versa. Note that it is
impossible to achieve any of the minimal lengths to vanish without the other
becoming infinitely large. We illustrate this in figure 1, where we plot $%
\Delta Y_{0}(\Delta X_{0})=\pm \theta \left\vert q^{2}-q^{-2}\right\vert
/(4\Delta X_{0})$ for a specific value of $\theta $ and various values of $q$%
. The two minimal areas indicated in the figure have the same size.

For a simultaneous $X,P_{x}$-measurement we compute similarly the minimal
momentum in the $X$-direction 
\begin{equation}
(\Delta P_{x})_{\text{min}}=\frac{\sqrt{(q^{2}-1)^{2}(\left\langle
Y\right\rangle ^{2}+\left\langle Y^{2}\right\rangle )\kappa _{3}^{2}\kappa
_{1}^{2}+\hbar \left\vert q^{4}-1\right\vert \kappa _{1}\kappa _{3}\kappa
_{5}^{2}+\left\langle X\right\rangle ^{2}(q^{2}-1)^{2}\kappa _{3}^{2}\kappa
_{5}^{2}}}{(q^{2}+1)\kappa _{1}\kappa _{5}},
\end{equation}
such that the corresponding absolute value turns out to be 
\begin{equation}
(\Delta P_{x})_{0}=2\kappa _{3}\frac{\sqrt{\left\vert q^{2}-1\right\vert }}{%
q^{2}+1}.
\end{equation}
There is no minimal length for $X$ in this case as we can tune $\Delta X$ to
be as small as we wish by enlarging $\Delta P_{x}$.

\noindent \includegraphics[width=14.0cm,height=10.0cm]{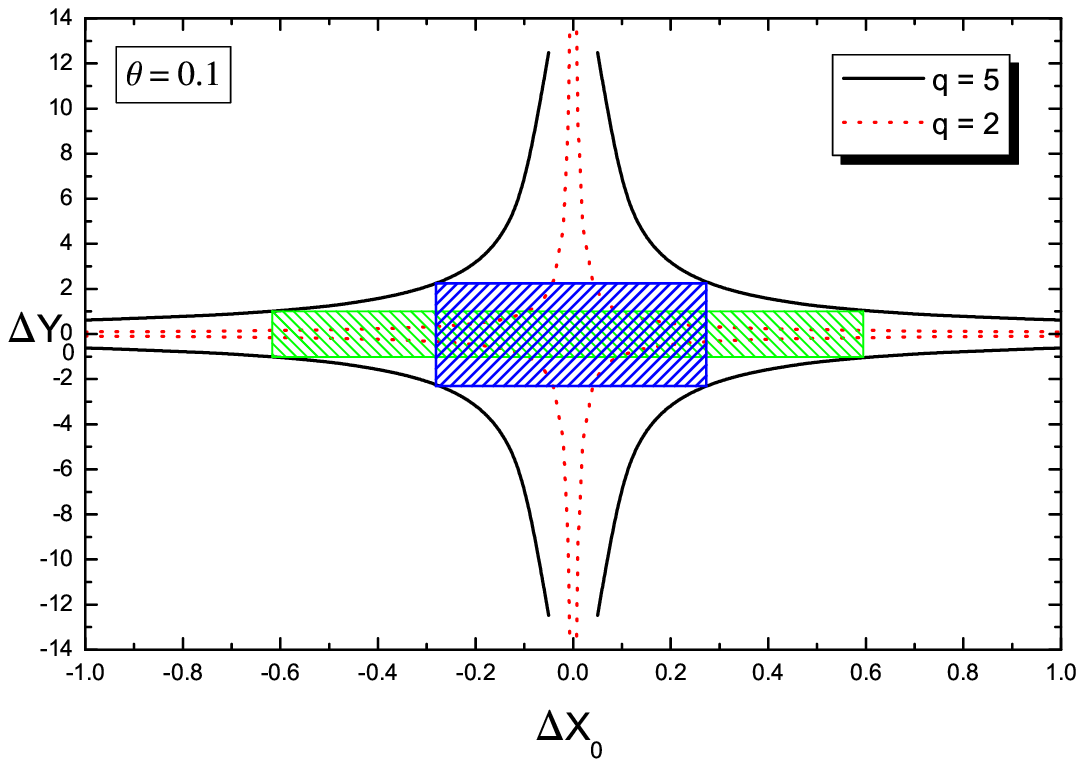}

\vspace{-1.0cm}

\noindent {\small Figure 1: Minimal areas in the XY-plane.}

Similarly we compute for a simultaneous $Y,P_{y}$-measurement the minimal
momentum in the $Y$-direction 
\begin{equation}
(\Delta P_{y})_{\text{min}}=\frac{\sqrt{(q^{2}-1)^{2}(\left\langle
X\right\rangle ^{2}+\left\langle X^{2}\right\rangle )\kappa _{7}^{2}\kappa
_{5}^{2}+\hbar \left\vert 1-q^{4}\right\vert \kappa _{5}\kappa _{7}\kappa
_{1}^{2}+\left\langle Y\right\rangle ^{2}(q^{2}-1)^{2}\kappa _{1}^{2}\kappa
_{7}^{2}}}{(q^{2}+1)\kappa _{1}\kappa _{5}},
\end{equation}%
with corresponding absolute value 
\begin{equation}
(\Delta P_{y})_{0}=2\kappa _{7}\frac{\sqrt{\left\vert q^{2}-1\right\vert }}{%
q^{2}+1}.
\end{equation}%
By the same reasoning as in the previous case there is also no minimal
length for $Y$ in this case as $\Delta Y$ can be taken to be as small as
desiredh by enlarging $\Delta P_{y}$.

The analysis for a simultaneous $P_{x},P_{y}$-measurement is less
straightforward due to the appearance of the angular momentum term. we first
note that 
\begin{eqnarray}
\left\vert \left\langle \lbrack P_{x},P_{y}]\right\rangle \right\vert &\geq
&\left\vert \frac{q^{2}-1}{q^{2}+1}\right\vert \left[ \left\vert \kappa
_{4}\kappa _{7}+\kappa _{3}\kappa _{8}\right\vert \left( \frac{\kappa _{7}}{%
\kappa _{8}\kappa _{1}^{2}}\left\langle X^{2}\right\rangle -\left\vert \frac{%
\kappa _{3}}{\kappa _{4}\kappa _{5}^{2}}\right\vert \left\langle
Y^{2}\right\rangle \right) \right. \\
&&\left. +\frac{\kappa _{8}}{\kappa _{4}}\left\langle P_{x}^{2}\right\rangle
+\frac{\kappa _{4}}{\kappa _{8}}\left\langle P_{y}^{2}\right\rangle -2\frac{%
\kappa _{4}\kappa _{7}}{\kappa _{1}\kappa _{8}}\left\vert \left\langle
YP_{x}\right\rangle \right\vert -2\frac{\kappa _{3}\kappa _{8}}{\kappa
_{4}\kappa _{5}}\left\vert \left\langle XP_{y}\right\rangle \right\vert %
\right] ,  \notag
\end{eqnarray}%
where for definiteness we assumed that $\kappa _{3}^{2}<\kappa _{4}^{2}$.
Using next the estimate $\left\vert \left\langle AB\right\rangle \right\vert
\leq \Delta A\Delta B+\left\vert \left\langle A\right\rangle \left\langle
B\right\rangle \right\vert $ we compute 
\begin{equation}
\Delta P_{x}\Delta P_{y}\geq \frac{1}{2}\left\vert \frac{q^{2}-1}{q^{2}+1}%
\right\vert \left[ \frac{\kappa _{8}}{\kappa _{4}}\Delta P_{x}^{2}+\frac{%
\kappa _{4}}{\kappa _{8}}\Delta P_{y}^{2}-2\left\vert \frac{\kappa
_{4}\kappa _{7}}{\kappa _{1}\kappa _{8}}\right\vert \Delta Y\Delta P_{x}-2%
\frac{\kappa _{3}\kappa _{8}}{\kappa _{4}\kappa _{5}}\Delta X\Delta
P_{y}+\lambda \right] ,  \label{PPP}
\end{equation}%
with 
\begin{eqnarray}
\lambda &=&\frac{\kappa _{8}}{\kappa _{4}}\left\langle P_{x}\right\rangle
^{2}+\frac{\kappa _{4}}{\kappa _{8}}\left\langle P_{y}\right\rangle
^{2}+\left\vert \kappa _{4}\kappa _{7}+\kappa _{3}\kappa _{8}\right\vert
\left( \frac{\kappa _{7}}{\kappa _{8}\kappa _{1}^{2}}\left\langle
X\right\rangle ^{2}-\left\vert \frac{\kappa _{3}}{\kappa _{4}\kappa _{5}^{2}}%
\right\vert \left\langle Y\right\rangle ^{2}\right) \\
&&-2\left\vert \frac{\kappa _{4}\kappa _{7}}{\kappa _{1}\kappa _{8}}%
\right\vert \left\vert \left\langle Y\right\rangle \left\langle
P_{x}\right\rangle \right\vert -2\frac{\kappa _{3}\kappa _{8}}{\kappa
_{4}\kappa _{5}}\left\vert \left\langle X\right\rangle \left\langle
P_{y}\right\rangle \right\vert .  \notag
\end{eqnarray}%
When varying the inequality (\ref{PPP}) in the same manner as the
expressions above we find 
\begin{eqnarray}
(\Delta P_{x})_{\text{min}} &=&-\frac{\left\vert q^{4}-1\right\vert }{4q^{2}}%
\frac{\kappa _{3}\kappa _{8}}{\kappa _{4}\kappa _{5}}\Delta X-\frac{\left(
q^{2}-1\right) ^{2}}{4q^{2}}\left\vert \frac{\kappa _{4}\kappa _{7}}{\kappa
_{1}\kappa _{8}}\right\vert \frac{\kappa _{4}}{\kappa _{8}}\Delta Y \\
&&\pm \frac{\left\vert q^{2}-q^{-2}\right\vert }{4}\sqrt{\frac{\kappa
_{3}^{2}\kappa _{8}^{2}\Delta X^{2}}{\kappa _{5}^{2}\kappa _{4}^{2}}+\frac{%
\kappa _{7}^{2}\kappa _{4}^{4}\text{$\Delta Y^{2}$}}{\kappa _{1}^{2}\kappa
_{8}^{4}}+\frac{2\left\vert \frac{\kappa _{4}\kappa _{7}}{\kappa _{1}\kappa
_{8}}\right\vert \kappa _{3}\Delta X\Delta Y}{\kappa _{5}\left\vert
q^{2}-1\right\vert (q^{2}+1)^{-1}}+\frac{4q^{2}\lambda \kappa _{4}}{\kappa
_{8}\left( q^{2}-1\right) ^{2}}}.  \notag
\end{eqnarray}%
and 
\begin{eqnarray}
(\Delta P_{y})_{\text{min}} &=&-\frac{\left( q^{2}-1\right) ^{2}}{4q^{2}}%
\frac{\kappa _{3}}{\kappa _{1}\kappa _{4}^{2}\kappa _{5}^{2}}\Delta X-\frac{%
\left\vert 1-q^{4}\right\vert }{4q^{2}}\left\vert \frac{\kappa _{4}\kappa
_{7}}{\kappa _{1}\kappa _{8}}\right\vert \frac{1}{\kappa _{1}\kappa
_{5}\kappa _{8}^{2}}\Delta Y \\
&&\pm \frac{\left\vert q^{2}-q^{-2}\right\vert }{4}\sqrt{\frac{\kappa
_{3}^{2}\kappa _{8}^{4}\Delta X^{2}}{\kappa _{4}^{4}\kappa _{5}^{2}}+\frac{%
\kappa _{4}^{2}\kappa _{7}^{2}\text{$\Delta Y^{2}$}}{\kappa _{1}^{2}\kappa
_{8}^{2}}+\frac{2\left\vert \frac{\kappa _{7}\kappa _{8}}{\kappa _{1}\kappa
_{4}}\right\vert \kappa _{3}\Delta X\Delta Y}{\kappa _{5}\left(
q^{2}-1\right) ^{2}\left\vert 1-q^{4}\right\vert ^{-1}}+\frac{4q^{2}\lambda
\kappa _{8}}{\kappa _{4}\left( q^{2}-1\right) ^{2}}}.  \notag
\end{eqnarray}%
We can minimize this expression further with a subsequent $X,Y$-measurement.
This is, however, a matter of interpretation if one would like to view
measurements as a pairwise succession or whether this should be considered
as a simultaneous measurement of four quantities. A further option would be
to exploit the explicit occurrence of the $L_{z}$-operator and take this
complication here as a hint that the angular momentum variables are possibly
a more natural set of variables. We leave this problem for future
investigations. Similar expressions are obtained for the choice $\kappa
_{3}^{2}>\kappa _{4}^{2}$.

\section{Conclusions}

We have demonstrated that dynamical noncommutative space-time relations will
inevitably lead to deformed oscillator algebras. Taking some well studied
oscillator algebras with the useful property that the entire Fock spaces
associated to them is explicitly constructable as a starting point, we
derived some very general commutation relations (\ref{XY})-(\ref{pxpy}) for
the dynamical variables. Since these relations are rather cumbersome, we
investigated some specific limits leading to simplified and more tractable
variants, whose properties can be discussed more transparently. All of these
special limits led to minimal lengths in the two dimensional space and
mostly to minimal areas which we have calculated explicitly (\ref{mina}).

There are some obvious further problems following from our considerations.
First of all it would be very interesting to explore the consequences of
taking different types of deformations as starting points and derive the
resulting dynamical commutation relations. Secondly it would be interesting
to consider explicit models on these type space-time structures and thirdly
but not last a generalization to three dimensional space would be highly
interesting.\medskip\ The latter will almost inevitably lead to minimal
volumes.

\noindent \textbf{Acknowledgments:} A.F. would like to thank the UGC Special
Assistance Programme in the Applied Mathematics Department of the University
of Calcutta and S.N. Bose National Centre for Basic Sciences for providing
infrastructure and financial support. Thanks for extremely kind hospitality
go to many members of these institutions, but especially to Bijan Bagchi and
Partha Guha for being tireless in this effort. L.G. is supported under the
grant of the National Research Foundation of South Africa.

\end{document}